\documentclass[reprint,superscriptaddress,amsmath,amssymb,prc, 
]{revtex4-2}
\UseRawInputEncoding
\usepackage{hyperref}
\usepackage{CJK}
\usepackage{graphicx}
\usepackage{booktabs}
\usepackage{bm}

\begin{document}
\begin{CJK*}{UTF8}{} 

\title{\protect Energy, strength, and alpha width measurements of $E_{\rm{c.m.}} = 1323$ and $1487$~keV resonances in $^{15}$N($\alpha,\gamma$)$^{19}$F}

\author{Ruoyu~Fang ({\CJKfamily{gbsn}方若愚})}
\email{rfang@nd.edu}
\affiliation{Department of Physics and Astronomy, University of Notre Dame, Notre Dame, Indiana 46556, USA}%
\affiliation{The Joint Institute for Nuclear Astrophysics - Center for the Evolution of Elements, Notre Dame, Indiana 46556, USA}%
\author{J.~G{\"o}rres}
\affiliation{Department of Physics and Astronomy, University of Notre Dame, Notre Dame, Indiana 46556, USA}%
\affiliation{The Joint Institute for Nuclear Astrophysics - Center for the Evolution of Elements, Notre Dame, Indiana 46556, USA}%
\author{R.J.~deBoer}
\affiliation{Department of Physics and Astronomy, University of Notre Dame, Notre Dame, Indiana 46556, USA}%
\affiliation{The Joint Institute for Nuclear Astrophysics - Center for the Evolution of Elements, Notre Dame, Indiana 46556, USA}%
\author{S.~Moylan}
\affiliation{Department of Physics and Astronomy, University of Notre Dame, Notre Dame, Indiana 46556, USA}%
\affiliation{The Joint Institute for Nuclear Astrophysics - Center for the Evolution of Elements, Notre Dame, Indiana 46556, USA}%
\author{A.~Sanchez}
\affiliation{Department of Physics and Astronomy, University of Notre Dame, Notre Dame, Indiana 46556, USA}%
\affiliation{The Joint Institute for Nuclear Astrophysics - Center for the Evolution of Elements, Notre Dame, Indiana 46556, USA}%
\author{T.L.~Bailey}
\affiliation{Department of Physics and Astronomy, University of Notre Dame, Notre Dame, Indiana 46556, USA}%
\affiliation{The Joint Institute for Nuclear Astrophysics - Center for the Evolution of Elements, Notre Dame, Indiana 46556, USA}%
\author{S.~Carmichael}
\affiliation{Department of Physics and Astronomy, University of Notre Dame, Notre Dame, Indiana 46556, USA}%
\affiliation{The Joint Institute for Nuclear Astrophysics - Center for the Evolution of Elements, Notre Dame, Indiana 46556, USA}%
\author{J.~Koros}
\affiliation{Department of Physics and Astronomy, University of Notre Dame, Notre Dame, Indiana 46556, USA}%
\affiliation{The Joint Institute for Nuclear Astrophysics - Center for the Evolution of Elements, Notre Dame, Indiana 46556, USA}%
\author{K.~Lee}
\affiliation{Department of Physics and Astronomy, University of Notre Dame, Notre Dame, Indiana 46556, USA}%
\affiliation{The Joint Institute for Nuclear Astrophysics - Center for the Evolution of Elements, Notre Dame, Indiana 46556, USA}%
\author{K.~Manukyan}
\affiliation{Department of Physics and Astronomy, University of Notre Dame, Notre Dame, Indiana 46556, USA}%
\author{M.~Matney}
\affiliation{Department of Physics and Astronomy, University of Notre Dame, Notre Dame, Indiana 46556, USA}%
\affiliation{The Joint Institute for Nuclear Astrophysics - Center for the Evolution of Elements, Notre Dame, Indiana 46556, USA}%
\author{J.P.~McDonaugh}
\affiliation{Department of Physics and Astronomy, University of Notre Dame, Notre Dame, Indiana 46556, USA}%
\affiliation{The Joint Institute for Nuclear Astrophysics - Center for the Evolution of Elements, Notre Dame, Indiana 46556, USA}%
\author{D.~Robertson}
\affiliation{Department of Physics and Astronomy, University of Notre Dame, Notre Dame, Indiana 46556, USA}%
\affiliation{The Joint Institute for Nuclear Astrophysics - Center for the Evolution of Elements, Notre Dame, Indiana 46556, USA}%
\author{J.~Rufino}
\affiliation{Department of Physics and Astronomy, University of Notre Dame, Notre Dame, Indiana 46556, USA}%
\affiliation{The Joint Institute for Nuclear Astrophysics - Center for the Evolution of Elements, Notre Dame, Indiana 46556, USA}%
\author{E.~Stech}
\affiliation{Department of Physics and Astronomy, University of Notre Dame, Notre Dame, Indiana 46556, USA}%
\affiliation{The Joint Institute for Nuclear Astrophysics - Center for the Evolution of Elements, Notre Dame, Indiana 46556, USA}%
\author{M.~Couder}
\email{mcouder@nd.edu}
\affiliation{Department of Physics and Astronomy, University of Notre Dame, Notre Dame, Indiana 46556, USA}%
\affiliation{The Joint Institute for Nuclear Astrophysics - Center for the Evolution of Elements, Notre Dame, Indiana 46556, USA}%

\date{\today}

\begin{abstract}
%
%
The $^{15}$N($\alpha,\gamma$)$^{19}$F reaction produces $^{19}$F in asymptotic giant branch (AGB) stars, where the low energy tails of two resonances at $E_{\rm{c.m.}} = 1323 \pm 2$ and $1487 \pm 1.7$~keV are estimated to contribute about 30\% of the total reaction rate in these environments.
%
%
However, recent measurements have shown discrepancies in the energies, the strengths, and the corresponding alpha widths of these two resonances, resulting in an increase in the systematic uncertainty of the extrapolated cross section to helium burning energies.
%
%
%
%
With this motivation, we have undertaken new measurements of the $^{15}$N$(\alpha,\gamma)^{19}$F at the University of Notre Dame Nuclear Science Laboratory. The setup consisted of an alpha particle beam impinged on a solid Ti$^{15}$N target with gamma-ray spectroscopy accomplished using a high purity germanium detector.
%
%
Using the Doppler corrected gamma-ray energies, we confirmed the lower resonance energy to be $1321.6 \pm 0.6$~keV and found a value for the higher one of $1479.4 \pm 0.6$~keV that is more consistent with those found from previous elastic scattering studies. We found that the resonance strengths for both were consistent with most values found in the literature, but a larger alpha width has been recommended for the $E_{\rm{c.m.}} = 1487$~keV resonance.
%
%
The larger alpha width suggests a reaction rate increase of about $15\%$ at temperatures $T < 0.1$~GK relevant to low mass AGB stars.
%
The impact of the increased reaction rate requires further investigations.
%
\end{abstract}

\maketitle
\end{CJK*}

\section{Introduction}
Fluorine is a key element in stellar evolution, geochemical, and biogeochemical systems~\cite{KOGA2018749},  yet the astrophysical origin of $^{19}$F is still unclear. The abundance of $^{19}$F in the universe is not fully explained by stellar models because of open questions and discrepancies in experimental nuclear physics inputs. Thus theoretical calculations of $^{19}$F production and destruction rates still have large uncertainties.

Several stellar environments, including core-collapse Supernovae~\cite{1988Natur.334...45W}, Wolf-Rayet stars~\cite{2000A&A...355..176M}, and AGB stars~\cite{jorissen_fluorine_1992} have been proposed to contribute to $^{19}$F production. The element's production has only been observed in AGB stars and the reported abundance was up to 30 times that of solar~\cite{jorissen_fluorine_1992}. However, subsequent studies have revised this number downward by up to a factor of six (see Refs.~\cite{abia_fluorine_2009, abia_fluorine_2010, abia_fluorine_2015}) because of corrections to the evaluation of the star spectroscopy~\cite{jonsson_chemical_2014}.

The abundance of $^{19}$F in AGB stars depends on its production and destruction rates. Refs.~\cite{lucatello_fluorine_2011, cristallo_effects_2014} provide detailed reviews on the different $^{19}$F reaction channels in AGB stars. The destruction of $^{19}$F primarily relies on the $^{19}$F$(p,\alpha)^{16}$O and $^{19}$F$(\alpha,p)^{22}$Ne reactions. The proton capture reaction has been studied extensively in recent years, reducing the reaction rate uncertainties significantly at relevant temperatures~\cite{la_cognata_fluorine_2011, lombardo_toward_2015, indelicato_2017, deboer_19F_2021, Zhang2022}. For the alpha capture reaction, the experimental efforts face the challenge of a higher Coulomb barrier. One direct measurement in the energy range $E_{\rm{c.m.}} = 0.66-1.6$ MeV has been reported by~\citet{ugalde_thermonuclear_2008}. The cross section was measured and an $R$-matrix fit was performed. The computed reaction rate at relevant temperatures has large uncertainties of about 50\% for most temperatures. Recently, a Trojan Horse Method (THM) measurement~\cite{pizzone_first_2017} suggested an increase of up to a factor of four in the reaction rate at astrophysical temperatures. This uncertainty in the reaction rate further complicates the understanding of $^{19}$F's destruction rate in AGB stars, especially in higher mass AGB stars where alpha capture reactions are more efficient because of higher temperatures~\cite{cristallo_effects_2014}.

Several reaction chains have been proposed as the production paths of $^{19}$F in AGB stars, with two leading. The first nuclear reaction chain $^{14}$N$(\alpha,\gamma)^{18}$F$(\beta^+)^{18}$O$(p,\alpha)^{15}$N$(\alpha,\gamma)^{19}$F was proposed by~\cite{jorissen_fluorine_1992}, however, subsequent AGB modeling calculations suggested the need for additional reaction chains to reproduce the observed abundance. \citet{forestini1992fluorine} proposed that a second nuclear reaction chain, $^{14}$N$(n,p)^{14}$C$(\alpha,\gamma)^{18}$O$(p,\alpha)^{15}$N$(\alpha,\gamma)^{19}$F, could also lead to the production of $^{19}$F in AGB stars. Both proposed reaction chains depend on $^{14}$N, the primary composition of the CNO ashes. 

The final stage of both reaction chains depends on the $^{15}$N($\alpha,\gamma$)$^{19}$F reaction, which is the main contributor to the production of $^{19}$F in AGB stars according to model calculations~\cite{cristallo_effects_2014}. The reaction rate at AGB temperatures mostly depends on alpha direct captures and narrow resonances, especially the $E_{\rm{c.m.}} = 364$~keV resonance with a corresponding $J^\pi$~=~7/2$^+$. This resonance has been studied only once through an indirect measurement, which resulted in a resonance strength uncertainty of about 100\%~\cite{DEOLIVEIRA1996231}.

In addition, the low energy tails of two resonances at $E_{\rm{c.m.}} = 1323$ and $1487$~keV can also contribute to the production of $^{19}$F because of their relatively large alpha widths. To calculate their contributions to the production of $^{19}$F, accurate knowledge of their resonance energies, alpha widths, and resonance strengths is needed. Recent measurements have shown discrepancies for these quantities. In order to investigate them, we present a new gamma-ray spectroscopy study of the $E_{\rm{c.m.}} = 1323$ and $1487$~keV resonances in the $^{15}$N$(\alpha,\gamma)^{19}$F reaction performed at the University of Notre Dame Nuclear Science Laboratory (NSL)~\cite{aprahamian_2014}.

\begin{figure*}
    \centering
    \includegraphics[width = \linewidth]{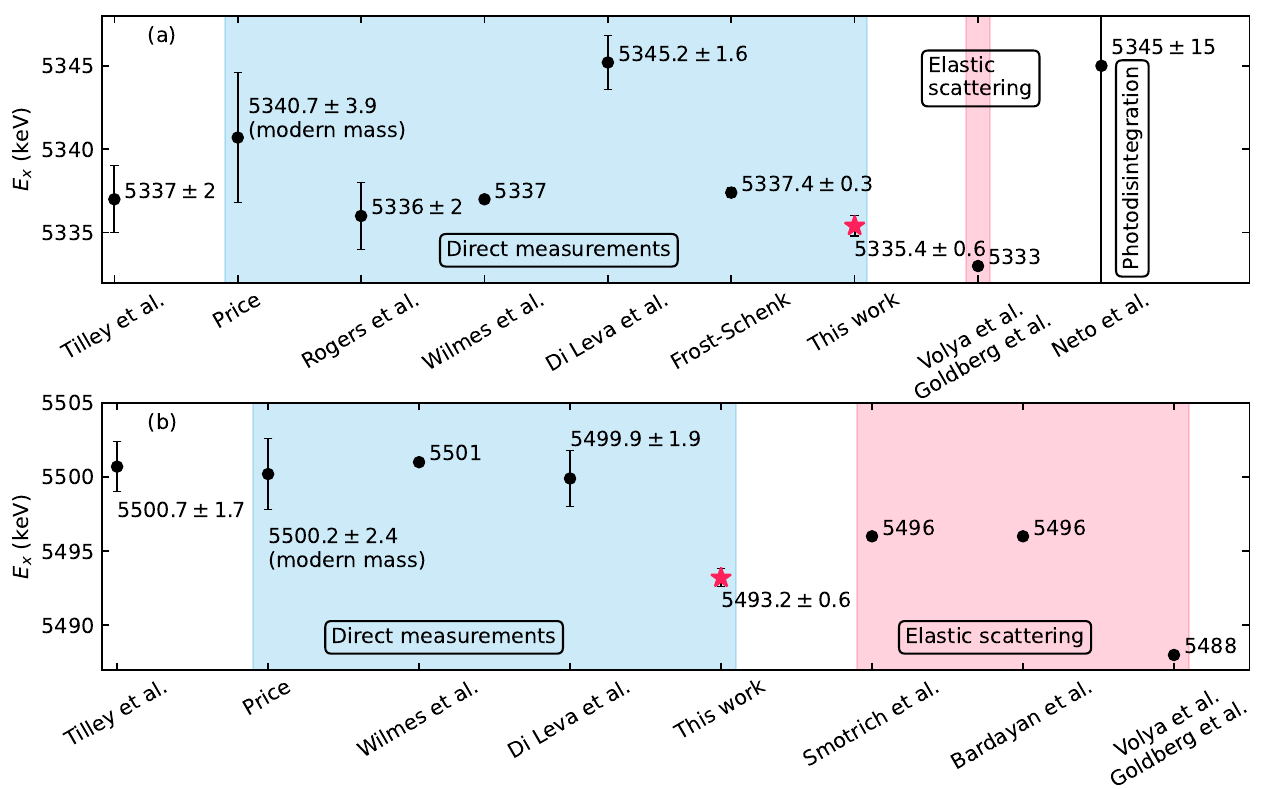}
    \caption{Status of this work and literature (a) on the $E_x = 5337$~keV excitation energy, (b) on the $E_x = 5501$~keV excitation energy. Different methods of studying these two excitation energies are labelled and shaded accordingly.}
    \label{fig:E_res_comparison}
\end{figure*}

The paper is structured as follows. Sec.~\ref{sec:lit_review} discusses the literature status on these two resonances. Sec.~\ref{sec:setup} provides a description of our experimental setup, methods employed, and details regarding the target properties. In Sec.~\ref{sec:results}, we discuss the analysis methods and results obtained from our measurement. We then compare our results with literature values and provide some discussions in Sec.~\ref{sec:discussion}. In Sec.~\ref{sec:rr}, we compare the astrophysical reaction rates to those given in~\citet{iliadis_charged-particle_2010}. Finally, in Sec.~\ref{sec:sum}, we summarize our results and the future directions this research suggests.

\section{Literature review}
\label{sec:lit_review}
\begin{figure*}
    \centering
    \includegraphics[width = \linewidth]{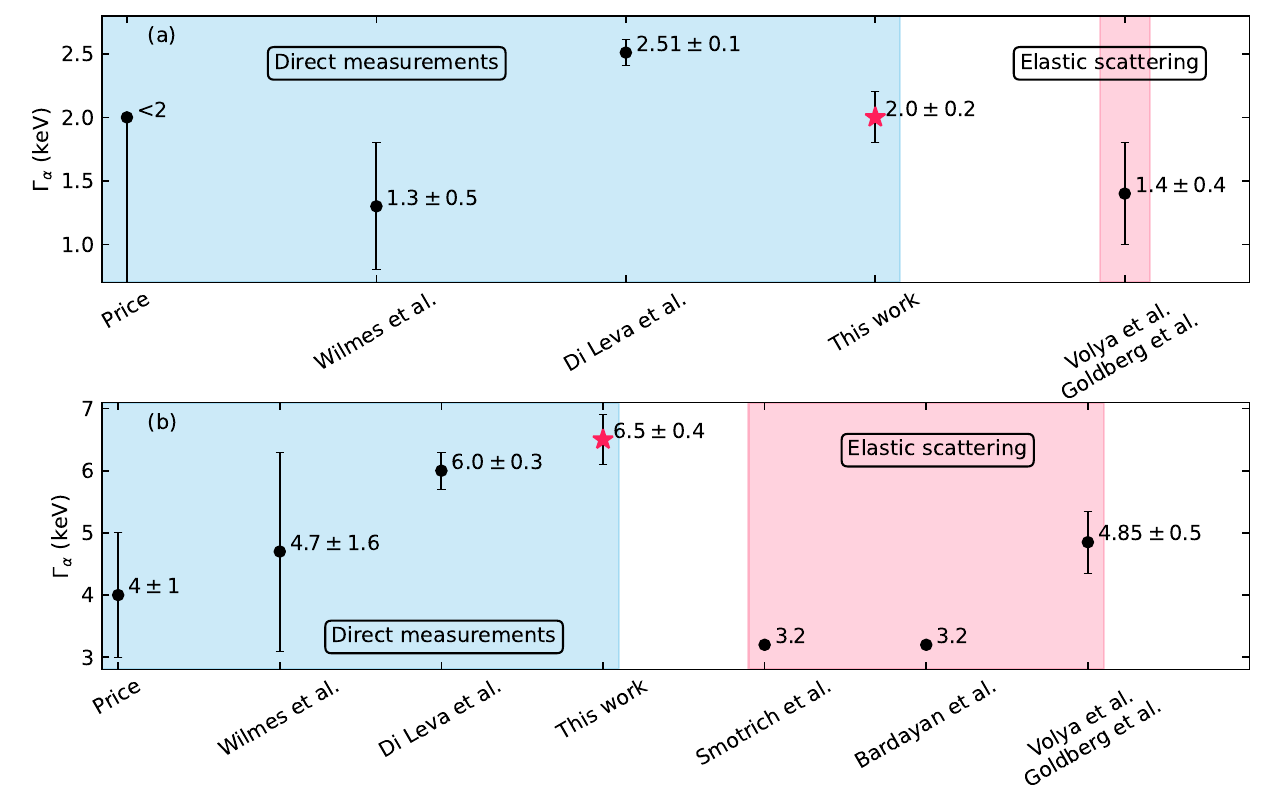}
    \caption{Status of this work and literature (a) on the $E_x = 5337$~keV alpha width. (b) on the $E_x = 5501$~keV alpha width. Different methods of studying these two alpha widths are labelled and shaded accordingly.}
    \label{fig:Gamma_comparison}
\end{figure*}

To avoid confusion, as inconsistent reference frames for the $E_{\rm{c.m.}} = 1323$ and $1487$~keV resonances have been used in the past, we will discuss these two resonances in the center of mass frame and in $^{19}$F excitation energies using masses from AME2020~\cite{wang_ame_2021}.

\subsection{Review of the study of the $J^{\pi} = 1/2^+$, $E_x = 5337$~keV level or $E_{\rm{c.m.}} = 1323$~keV resonance}
\label{sec:5337_review}
\citet{Price57} first directly measured this resonance in 1957 in forward kinematics with thick tantalum nitride solid targets, which were made by heating tantalum foils in an atmosphere of $^{15}$N enriched ammonium nitrate. The emitted gamma-rays were detected by NaI(Ti) crystals. Using nuclear mass values available at the time~\cite{li_nuclear_1952}, \citet{Price57} determined the alpha beam energy corresponding to the resonance from the thick-target yield and converted it to an excitation energy of $5313 \pm 3.9$~keV. When revised using modern mass tables~\cite{wang_ame_2021}, the excitation energy becomes $5341 \pm 3.9$~keV.

This excitation energy was later measured again by \citet{rogers_1972}. Instead of relying on beam energy, they measured the emitted gamma-ray energies and determined that the corresponding excitation energy was at $E_x = 5336 \pm 2$~keV, a value consistent with that given in \citet{Price57}. This excitation energy was further solidified by experiments populating the state with channels other than $(\alpha,\gamma)$ and the excitation energy reported in the compilation \cite{tilley_energy_1995} was determined to be $5337 \pm 2$~keV (see Table~19.9 in the compilation~\cite{tilley_energy_1995} and references therein).

However, this excitation energy has recently been challenged by \citet{2017PhRvC..95d5803D}, who proposed a larger value of $E_x = 5345.2 \pm 1.6$~keV (deduced from the reported center of mass energy) from their inverse kinematics studies with the European Recoil separator for Nuclear Astrophysics (ERNA). In addition, using the Detector Of Recoils And Gamma-rays of Nuclear reactions (DRAGON) at TRIUMF, Canada, \citet{Frost-Schenk2020} also investigated this resonance in inverse kinematics and reported the resonance to be at $E_x = 5337.4 \pm 0.3$ (deduced from the reported center of mass energy) although the work has not been peer-reviewed. Most recently, \citet{neto_2023} performed the time-inverse photo-dissociation measurement $^{19}$F($\gamma,\alpha$)$^{15}$N. They reported an excitation energy of $5345 \pm 15$~keV, adding more uncertainties to the exact excitation energy of this level. Therefore, one of the motivations of this work is to investigate the excitation energy of this level.

Resonance strength studies were pioneered by \citet{Aitken_1970} followed by \citet{Dixon_Storey_1971}. They first reported a resonance strength of $\omega \gamma = 1.30 \pm 0.20$~eV, making a measurement relative to the $E_{\alpha} = 1532$~keV resonance in the $^{14}$N($\alpha,\gamma$)$^{18}$F reaction~\cite{parker_n_1968}. \citet{Dixon_Storey_1971} reported three different resonance strengths in their study. First, they made the same type of relative resonance strength measurements and reported a resonance strength of $\omega \gamma = 1.63\pm 0.20$~eV. They then also did a relative measurement to the $E_{p} = 898$~keV resonance in the $^{15}$N$(p,\alpha_1 \gamma)^{12}$C reaction and found $\omega \gamma = 1.72 \pm 0.20$~eV. Lastly, they reported an absolute measurement finding $\omega\gamma = 1.5 \pm 0.3$~eV. Their recommended resonance strength was the weighted average of all three methods $\omega \gamma = 1.64 \pm 0.16$~eV, which has been reported in the latest compilation~\cite{tilley_energy_1995}. However, a re-evaluation of this value is necessary due to updates in the resonance strengths of reactions that \citet{Dixon_Storey_1971} relied on for their relative measurements. See Appendix~\ref{sec:re_evaluation_of_wy_at_5337} for details.

More recently, \citet{Wilmes} and \citet{2017PhRvC..95d5803D} reported measurements of the $E_{\rm{c.m.}} = 1323$~keV resonance strength that are compatible with \citet{Dixon_Storey_1971}. On the other hand, \citet{Frost-Schenk2020} reports a resonance strength of $\omega \gamma = 0.92 \pm 0.11$~eV, which is 40\% lower than those in \cite{Dixon_Storey_1971}, \cite{Wilmes}, and \cite{2017PhRvC..95d5803D}. The discrepancy needs to be investigated further because the strength of this resonance is used to normalize the strengths of 17 others in the $^{15}$N$(\alpha, \gamma)^{19}$F reaction at higher energy \cite{rogers_1972, dixon_storey_1977}. Any change to the $E_{\rm{c.m.}} = 1323$~keV resonance strength would therefore have a significant impact on the reaction rate over a wide temperature range. A summary of the values from the literature for this level has been presented in Figs.~\ref{fig:E_res_comparison}(a) and ~\ref{fig:wy_comparison}(a).

\subsection{Review of the study of $J^{\pi} = 3/2^+$, $E_x = 5501$~keV level or $E_{\rm{c.m.}} = 1487$~keV resonance}
\label{sec:5501 review}

Direct $(\alpha, \gamma)$ studies of the $E_x = 5501$~keV level are limited in the literature. \citet{Price57} first determined its energy to be $E_x = 5500 \pm 2.4$~keV (revised using modern mass tables) in the same measurement campaign where the lower energy level was investigated. The next direct energy measurement of this level was made by \citet{2017PhRvC..95d5803D} in inverse kinematics, which is consistent with that of \citet{Price57}.

Properties of this state have also been investigated using elastic scattering. \citet{smotrich_elastic_1961}'s study found this level to have an excitation energy of $5475$~keV. However due to change in nuclear masses, revisions have become necessary, motivating reanalysis of part of the original data of their work, at $\theta_{\rm{c.m.}} = 169.1 ^{\circ}$ and excitation energy up to 7300~keV, by \citet{bardayan_2005}. They concluded that the excitation energy should be $5496$~keV, suggesting a lower value for this state. Using elastic scattering and gamma-ray spectroscopy studies, \citet{Wilmes} claimed that they confirmed the excitation energy of this state as well, which is the value adopted by the compilation \cite{tilley_energy_1995}. However, they did not provide the data used to reach this conclusion and the uncertainties were not discussed in detail. More recently, \citet{volya_2022} and \citet{Goldberg2022} again reanalyzed the data from \citet{smotrich_elastic_1961}, where they included the data at all measured angles and over the complete energy region up to 8330~keV excitation energy. In addition, they also measured the excitation functions of the elastic scattering of several low lying and broad resonances including the corresponding $E_x = 5501$~keV state. They concluded that the excitation energy of this level is $5488$~keV, indicating that a lower energy is more likely.

To date, only two publications discuss the corresponding resonance strength of this level. One is the inverse kinematics measurement of \citet{2017PhRvC..95d5803D} and the other is that of \citet{Wilmes}. Yet, these resonance strengths are inconsistent with each other. It is also worth mentioning that a third value is proposed in the compilation \cite{tilley_energy_1995}, but it is a value deduced by Rogers from the resonance strength corresponding to the $E_x = 5337$~keV level from Ref.~\cite{Dixon_Storey_1971} through a private communication (see Table 19.7 in Ref.~\cite{ajzenberg_energy_1972}). This calculated value has a 25\% uncertainty, which makes it compatible with the published values in \citet{Wilmes} and \citet{2017PhRvC..95d5803D}. A summary of the literature for this level is presented in Figs.~\ref{fig:E_res_comparison}(b) and ~\ref{fig:wy_comparison}(b).

\section{Experimental Setup and Procedures} \label{sec:setup}
The experiment discussed here aims to address the literature discrepancies in the energies (see Fig.~\ref{fig:E_res_comparison}) and strengths of the $E_{\rm{c.m.}} = 1323$ and $1487$~keV resonances. Additionally, this work investigates discrepancies in the total widths (dominated by the alpha width) as shown in Fig.~\ref{fig:Gamma_comparison}, which may impact the reaction rates at temperatures relevant to AGB stars by up to 15\%~\cite{2017PhRvC..95d5803D}.

The experiment was performed at the NSL using the 5~MV Stable ion Accelerator for Nuclear Astrophysics (Sta. ANA). A singly ionized $^{4}$He$^+$ beam was produced over a laboratory energy range from $E_{\alpha}$ = 1.6 to 1.9~MeV with typical beam intensities of $\approx 10$~$\textrm{e}\mu$A on a Ti$^{15}$N target with a thick Ta backing. The energy uncertainty of the beam was measured to be better than 1~keV using the well-known, narrow, $^{27}$Al$(p,\gamma)^{28}$Si resonance at $E_p~=~992$~keV~\cite{brindhaban_accelerator_992res_1994}.

The experimental setup is identical to that of Fig.~4(a) in \citet{frentz_14Npg_2022} except that no lead shielding was used in this experiment. The targets were mounted on a target holder tilted at 45$^{\circ}$ with respect to the beam axis. To keep the targets from overheating under beam bombardment, the target backings were water cooled. A copper tube cold trap, biased to -400~V and cooled with liquid nitrogen, was installed in front of the target to limit carbon build-up and suppress secondary electrons from the target. The position of the beam at the target location was defined by a pair of vertical and horizontal slits. Moreover, the scintillation light emitted by the interaction of the beam and the target was observed continuously to monitor the beam position.

Gamma-rays were detected using a high purity germanium (HPGe) detector with a relative efficiency of 104\%. The HPGe detector was mounted on an electrically isolated sliding platform, allowing for convenient adjustment of the distance between the detector and the target. The majority of the measurements were conducted at a close distance of 4.4~cm from the target. In addition, a few measurements were performed at a farther distance of 20~cm to evaluate and correct for summing effects and the relative efficiency between the close and far distance setups. To mitigate angular distribution effects, the detector was positioned at an angle of 55$^{\circ}$ relative to the beam axis, as this angle corresponds to the minimum of the second-order Legendre polynomial~\cite{Devons1957}.

\subsection{Target}
\label{sec:target}
The Ti$^{15}$N target was fabricated at the Forschungszentrum Karlsruhe through reactive sputtering of Ti in a 99.5\% enriched $^{15}$N environment. \citet{leblanc_stoichiometry_2010} verified the target's stoichiomtery to be within a tolerance of $\leq 2\%$ compared to the nominal stoichiometry of 1:1. 

The energy loss of the beam in the target at $E_{\rm{c.m.}} = 1323$~keV was deduced from the excitation function (see Section \ref{sec:res_strength}) to be $11.7 \pm 0.7$~keV in the center of mass frame using the full width at half maximum (FWHM) from the thick target yield. The stability of the target was verified by checking the yield of the $E_{\rm{c.m.}} = 1323$~keV resonance at various times throughout the experiment. No target material loss was observed during the close distance measurements. However, a 23\% reduction in Ti$^{15}$N content was observed after depositing 0.7~C of beam from the longer duration, far distance, measurements at $E_{\rm{c.m.}} = 1323$~keV. In addition, the uncertainty in the alpha width of the $E_{\rm{c.m.}} = 1487$~keV resonance prevented the use of the FWHM method to deduce the energy loss for this resonance. For this reason, the energy loss of the beam in the target for the $E_{\rm{c.m.}} = 1487$~keV resonance measurement was determined to be $8.5 \pm 0.7$~keV in the center of mass frame by scaling the beam energy loss determined from the $E_{\rm{c.m.}} = 1323$~keV excitation function to the corresponding stopping power as a function of beam energy \cite{ziegler_srim_2010} and accounting for the target material loss. No additional target degradation was observed for the $E_{\rm{c.m.}} = 1487$~keV close distance excitation function measurements. 

\subsection{HPGe Detector} \label{sec:hpge_setup}
The energy calibration of the detector and the determination of its absolute efficiency were carried out using a calibrated $^{60}$Co source and the $^{27}$Al$(p,\gamma)^{28}$Si resonance at $E_p = 992$~keV. The branching ratios, angular distributions, and the absolute yield of $(1.08 \pm 0.06) \times 10^{-9}$ gamma-rays (1779~keV) per incident proton were well known for this resonance~\cite{antilla1977}. 

The detector energy calibration was characterized with a few gamma-ray transitions from the populated excited state of $^{28}$Si. To accurately determine the resonance energy of the $^{15}$N($\alpha,\gamma$)$^{19}$F reaction, a precise energy calibration of the HPGe detector was necessary. Corrections for the Doppler shift and recoil shift in the measured gamma-ray energies were considered and the calibration was thus determined to better than $1$~keV between $E_\gamma = 1.2$ and $10.8$ MeV.

\begin{figure}
    \centering
    \includegraphics[width=\linewidth]{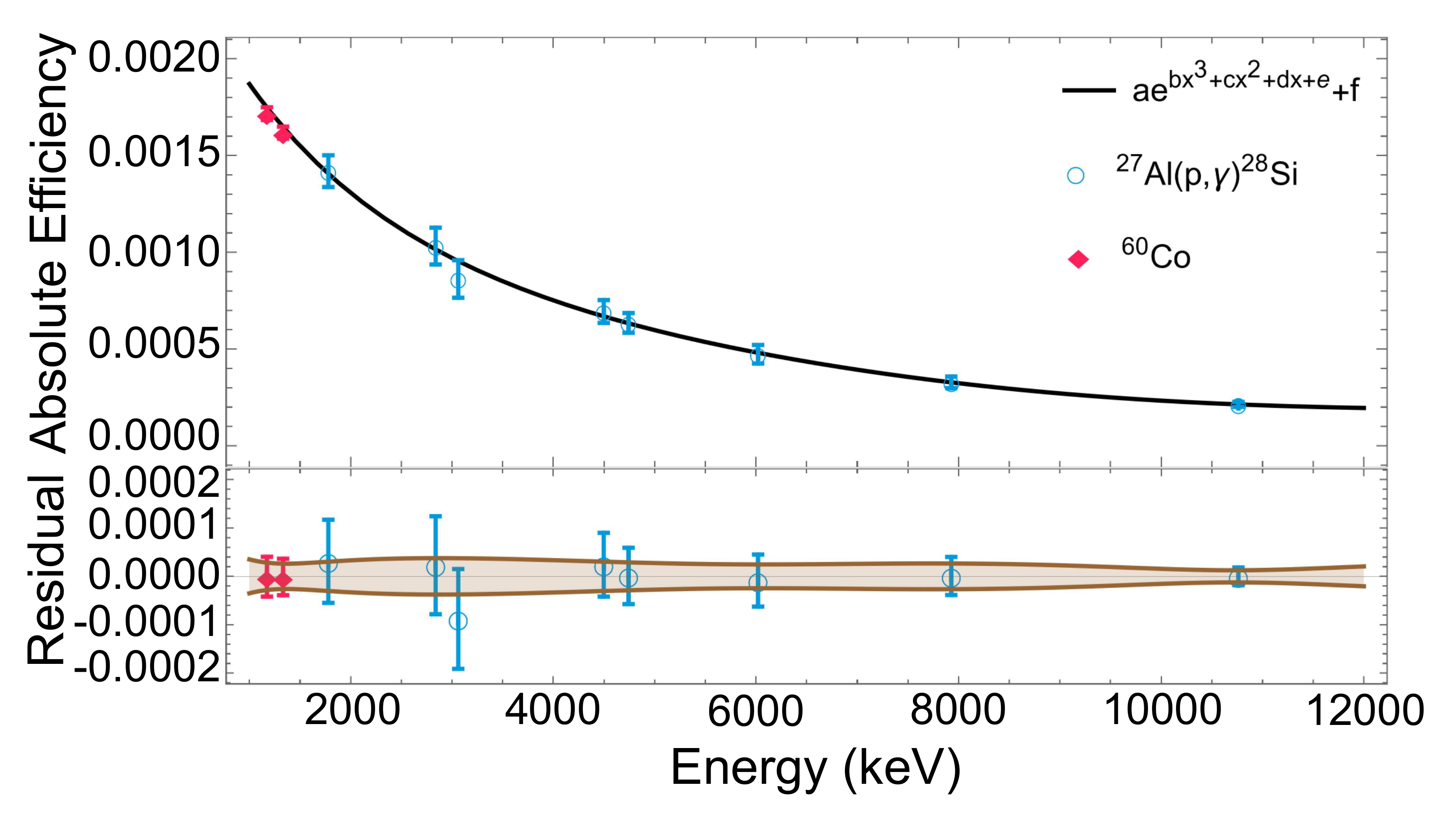}
    \caption{Absolute efficiency curve for the HPGe detector. The measured photopeak efficiencies were fitted with an exponential function indicated by the solid line. The 1$\sigma$ confidence band is shown as the shaded region in the residual plot.}
    \label{fig:gamma_efficiency}
\end{figure}

The absolute efficiency curve for the HPGe detector was obtained as shown in Fig.~\ref{fig:gamma_efficiency}. The efficiency at higher energies was normalized to the efficiency of the 1779~keV gamma-ray from the $^{27}$Al$(p,\gamma)^{28}$Si resonance. The low-energy end was obtained from a calibrated $^{60}$Co source. The uncertainty in the absolute efficiency was dominated by the yield uncertainty of the $^{27}$Al$(p,\gamma)^{28}$Si resonance, which was known to better than $\pm 7\%$ in the energy region of interest.

\section{Results} \label{sec:results}
\subsection{Resonance Energy} 

\begin{figure}
    \centering
    \includegraphics[width = \linewidth]{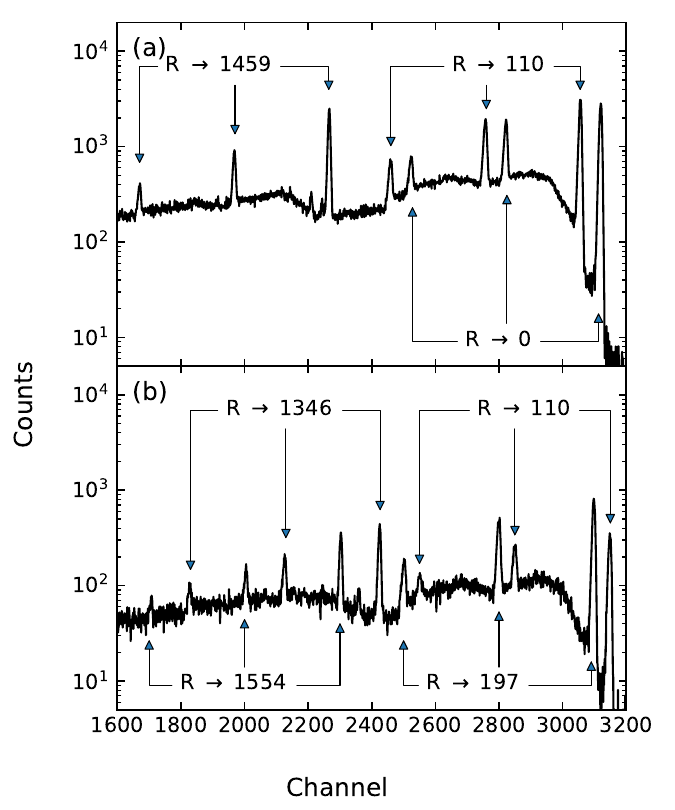}
    \caption{The energy spectrum measured by the HPGe detector when populating the $E_x = 5337$~keV state (panel (a)), and the $E_x = 5501$~keV state (panel (b)) using the $^{15}$N$(\alpha,\gamma)^{19}$F reaction. Transitions, including the single and double escape peaks to different final states, are labeled in each panel.}
    \label{fig:gamma_spec_stacked}
\end{figure}

The excitation functions over the $E_{\rm{c.m.}} = 1323$ and $1487$~keV resonances in the $^{15}$N($\alpha,\gamma$)$^{19}$F reaction were measured. Fig.~\ref{fig:gamma_spec_stacked} shows the gamma-ray spectrum for both resonances. The measured gamma-ray energy $E_{\gamma,m}$ at an angle $\theta$ relative to the beam axis is related to the excitation energy by (adapted from \cite{Iliadis2015})
\begin{equation}
\begin{split}
    E_{\gamma,m} =& E_x - E_f \\ 
    & + 4.63367 \times 10^{-2} \frac{\sqrt{m_a (E_x-Q) (m_a+m_A)/m_A}}{m_B} \\
    & \times (E_x-E_f) \rm{cos} \theta \\
    & - 5.36772 \times 10^{-4} \frac{(E_x-E_f)^2}{m_B},
\end{split}
    \label{Eq: Doppler_modified}
\end{equation}
where $E_f$ is the final state energy, $Q$ is the $Q$-value of the reaction and $m_a, m_A, m_B$ are masses of the projectile, target and product nuclei, respectively. Here, all energies are in units of MeV and the masses are in units of u.

Using the gamma transitions to the Ground State (G.S.), $E_x = 109.894$~keV state, and $E_x = 1458.7$~keV state~\cite{tilley_energy_1995} and correcting the measured gamma-ray energies for the Doppler shift as described by Eq.~(\ref{Eq: Doppler_modified}), three excitation energies are found and shown in Table~\ref{table::15N_1323_doppler_table}. The weighted average of those three values results in an excitation energy of $E_x = 5335.4 \pm 0.6$~keV or $E_{\rm{c.m.}} = 1321.6 \pm 0.6$~keV for this state.

\begin{table}
\caption{\label{table::15N_1323_doppler_table} Doppler shift corrected photopeak energies for the $E_{\rm{c.m.}} = 1323$~keV resonance in the $^{15}$N$(\alpha,\gamma)^{19}$F reaction. All energies are in units of~keV.}
\begin{ruledtabular}
\begin{tabular}{cccc}
$E_{\gamma}$\footnote{This work.} & $E_f$\footnote{Ref.~\cite{tilley_energy_1995}}& $E_x$\footnotemark[1] & $E_{\rm{c.m.}}$\footnotemark[1]\\
\midrule
5335.3(9)  & G.S.		&  5335.3(9)	&	1321.5(9)\\
5225.9(9)  & 109.894(5)&  5335.8(11)	&	1322.0(11)\\
3876.4(9)  & 1458.7(3)	&  5335.1(10)	&   1321.3(10)\\
\midrule
Weighted avg. & - & 5335.4(6) & 1321.6(6)\\
\end{tabular}
\end{ruledtabular}
\end{table}

Using the same approach for the $E_{\rm{c.m.}} = 1487$~keV resonance, the energy for each transition could be determined individually. The results for each transition from the populated $E_{\rm{c.m.}} = 1487$~keV resonance are presented in Table~\ref{table::15N_1487_doppler_table}. The weighted average of the corresponding excitation energy deduced from the four transitions is $E_x = 5493.2 \pm 0.6$~keV ($E_{\rm{c.m.}} = 1479.4 \pm 0.6$~keV).

\begin{table}
\caption{\label{table::15N_1487_doppler_table} Doppler shift corrected photopeak energies for the $E_{\rm{c.m.}} = 1487$~keV resonance in the $^{15}$N$(\alpha,\gamma)^{19}$F reaction. All energies are in units of~keV.}
\begin{ruledtabular}
\begin{tabular}{cccc}
$E_{\gamma}$\footnote{This work.} & $E_f$\footnote{Ref.~\cite{tilley_energy_1995}} & $E_x$\footnotemark[1] & $E_{\rm{c.m.}}$\footnotemark[1]\\
\midrule
5383.7(22)	& 109.894(5)	& 5493.6(22)	&	1479.8(22)	\\
5296.6(9)	& 197.143(4)	& 5493.7(11)	&	1479.9(11)	\\
4147.1(10)	& 1345.67(13)	& 5492.8(11)	&	1479.0(11)	\\
3938.7(13)	& 1554.038(9)	& 5492.8(14)	&	1479.0(14)	\\
\midrule
Weighted avg. & - & 5493.2(6) & 1479.4(6)\\
\end{tabular}
\end{ruledtabular}
\end{table}

\subsection{Resonance Strength}
\label{sec:res_strength}
The general expression for the experimental yield is \cite{Iliadis2015}
\begin{equation}
\begin{split}
    Y(E_0) = & \int_{E_0 - \Delta E}^{E_0}dE' \int_{E_i = 0}^{E_i} dE_i \\
    & \cdot \int_{E = 0}^{E_i} \frac{\sigma(E)}{\epsilon(E)}g(E_0-E_i)f(E_i-E, E')dE,
\end{split}
\label{eq:general_yield}
\end{equation}
for a beam of mean energy $E_0$, an energy distribution $g(E_0- E_i)$, stopping power $\epsilon(E)$, energy loss in the target $\Delta E$, and energy loss and straggling described by $f(E_i-E, E')$. 

Assuming the resonance cross section follows the Breit-Wigner distribution, and that the effective stopping power $\epsilon_{\rm{eff}}$, the de Broglie wavelength $\lambda_r$, and the corresponding partial widths $\Gamma_i$ of the resonance can be treated as independent of the energy over the resonance width, the experimental yield can be expressed as
\begin{equation}
    \begin{split}
        Y(E_0) =& \frac{\lambda_r^2}{2\pi}\frac{\omega \gamma}{\epsilon_{\rm{eff}}} \left[ \rm{arctan}(\frac{E_0-E_{r}}{\sqrt{\Gamma^2+\Delta_{\rm{beam}}^2}/2}) \right.\\
        & \left. -\rm{arctan}(\frac{E_0-E_r-\Delta E}{\sqrt{\Gamma^2+\Delta_{\rm{beam}}^2 + \Theta_{\rm{target}}^2}/2}) \right],
    \end{split}
    \label{eq:yield_arctan}
\end{equation}
where $\Delta E$, $\Delta_{\rm{beam}}$ and $\Theta_{\rm{target}}$ represent the beam energy loss in the target, beam energy resolution, and target inhomogenity effect, respectively. The effective stopping power in center of mass frame is \cite{Iliadis2015}
\begin{equation}
    \epsilon_{\rm{eff}} = \frac{m_{\rm{^{15}N}}}{m_{\rm{^{4}He}} + m_{\rm{^{15}N}}} \bigl[ \epsilon_{\rm{^{15}N}} + (\frac{N_{\rm{Ti}}}{N_{\rm{^{15}N}}}) \epsilon_{\rm{Ti}} \bigr],
    \label{eq:effective_stopping_power}
\end{equation}
where $m_{\rm{^{4}He}}$ and $m_{\rm{^{15}N}}$ are the masses of the beam and active target nuclei in units of u, $N_{\rm{Ti}}/N_{\rm{^{15}N}}$ is the stoichiometry of the target, and $\epsilon_{\rm{^{15}N}}$ and $\epsilon_{\rm{Ti}}$ are the stopping powers of the beam in $^{15}$N and Ti respectively. The stopping powers were obtained from the computer code SRIM \cite{ziegler_srim_2010}. At the two resonance energies under study in this work, the uncertainties of these stopping powers are 5\% and 4\% for $^{15}$N and Ti, respectively \cite{montanari_iaea_2017}.

For the analysis of the $E_{\rm{c.m.}} = 1323$~keV resonance, where the populated level has a $J^\pi$ = 1/2$^+$, the emitted gamma-rays are isotropic. However, the gamma-ray emission of the $E_{\rm{c.m.}} = 1487$~keV resonance is not expected to be isotropic because this level has a $J^\pi = 3/2^+$. Nevertheless, the HPGe detector location at 55$^{\circ}$ strongly reduces the dependence on higher order Legendre polynomial contributions to the gamma-ray yield~\cite{Iliadis2015}. \citet{Price57} has also demonstrated that the fourth order Legendre polynomial is negligible. Therefore, angular distribution corrections are negligible for this resonance as well with the present setup.

\begin{figure}
    \centering
    \includegraphics[width = \linewidth]{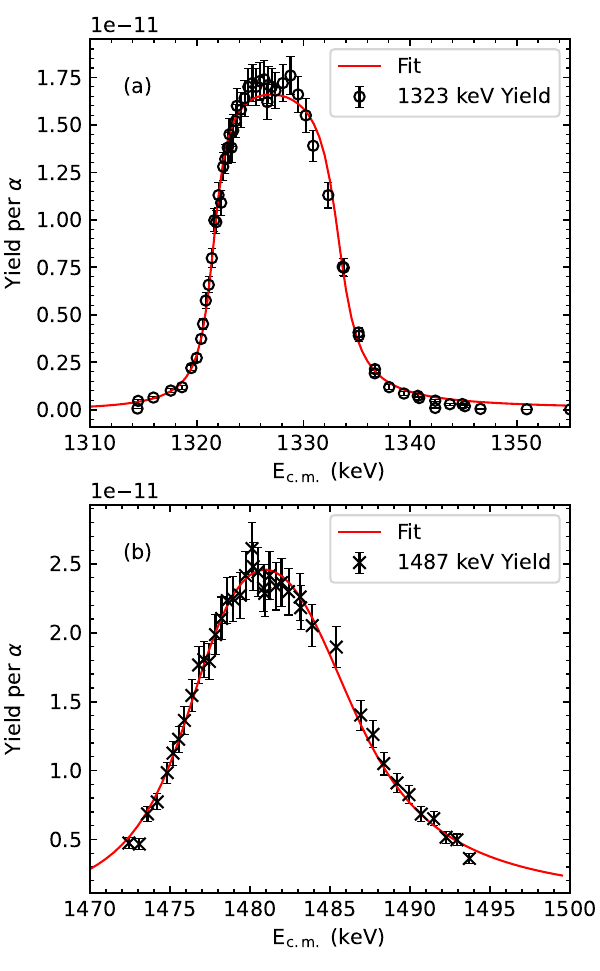}
    \caption{Yield per incident $\alpha$ particle for the $E_{\rm{c.m.}} = 1323$ and $E_{\rm{c.m.}} = 1487$~keV resonances, panel (a) and (b) respectively, obtained using the close geometry setup described in Sec.~\ref{sec:setup}. The red lines indicate the fits to the data using Eq.~(\ref{eq:yield_arctan}).}
    \label{fig:yield_arctan}
\end{figure}

The excitation functions for both resonances were fitted with Eq.~(\ref{eq:yield_arctan}) using the Least Square Fit (LSF) method as shown in Fig.~\ref{fig:yield_arctan}. Parameters $\omega\gamma$ (Table~\ref{table::res_strength_results}), $\Gamma$ (Fig.~\ref{fig:Gamma_comparison}), and $\Theta_{\rm{target}}$ were extracted from the fitting results. The reduced $\chi^2$ of each fit were $0.93$ and $1.3$ for the lower and higher energy resonance, respectively. 

In addition to fitting the experimental yield to Eq.~(\ref{eq:yield_arctan}), two other methods were used to deduce the resonance strength. First, the resonance strength can also be deduced from the maximum yield using \cite{Iliadis2015}
\begin{equation}
    Y(E)_{max} = \frac{\lambda_r^2}{\pi} \frac{\omega \gamma}{\epsilon_r} \rm{arctan}\frac{\Delta E}{\Gamma}.
    \label{eq: yield_max}
\end{equation}
With this method, we obtained resonance strengths of $\omega \gamma_{1323~\rm{keV}}=1.62 \pm 0.12$~eV and $\omega \gamma_{1487~\rm{keV}}=3.80 \pm 0.32$~eV. The uncertainty is dominated by the detector's efficiency calibration, which originated from the yield uncertainty of the $^{27}$Al$(p,\gamma)^{28}$Si resonance used for calibration. 

Second, we also used the ``area under the yield curve" method. The area is linked to the resonance strength through numerically evaluating Eq.~(\ref{eq:general_yield}) \cite{Iliadis2015}
\begin{equation}
\begin{split}
    A_Y &=  \int_{E_0 = 0}^{\infty} Y(E_0) dE_0 \\
        &= \frac{\Delta E}{\epsilon_r} \frac{\lambda_r^2}{2} \omega \gamma,
\end{split}
\end{equation}
assuming a Breit-Wigner cross section with constant stopping power, partial widths and de Broglie wavelength over the width of the resonance. Table~\ref{table::res_strength_results} provides a summary of resonance strengths obtained by the three methods.

\begin{table}[h!]
\caption{\label{table::res_strength_results} Strengths of the $E_{\rm{c.m.}} = 1323$ and $1487$ keV resonances in the $^{15}$N($\alpha,\gamma$)$^{19}$F reaction obtained by three methods. All strengths are in units of~eV.}
\begin{ruledtabular}
\begin{tabular}{ccccc}
 & Maximum & Yield curve & Yield curve &Weighted\\
 & yield & fitting & integration & avg.\\
\midrule
$\omega \gamma_{1323 \rm{~keV}}$ & $1.62 \pm 0.13$ & $1.63 \pm 0.11$ & $1.60 \pm 0.11$ & $1.61 \pm 0.07$\\
$\omega \gamma_{1487 \rm{~keV}}$ & $3.80 \pm 0.32$ & $4.19 \pm 0.38$ & $4.13 \pm 0.37$ & $4.01 \pm 0.20$\\
\end{tabular}
\end{ruledtabular}
\end{table}

\section{\label{sec:discussion} Discussion}
\begin{figure*}
    \centering
    \includegraphics[width = \linewidth]{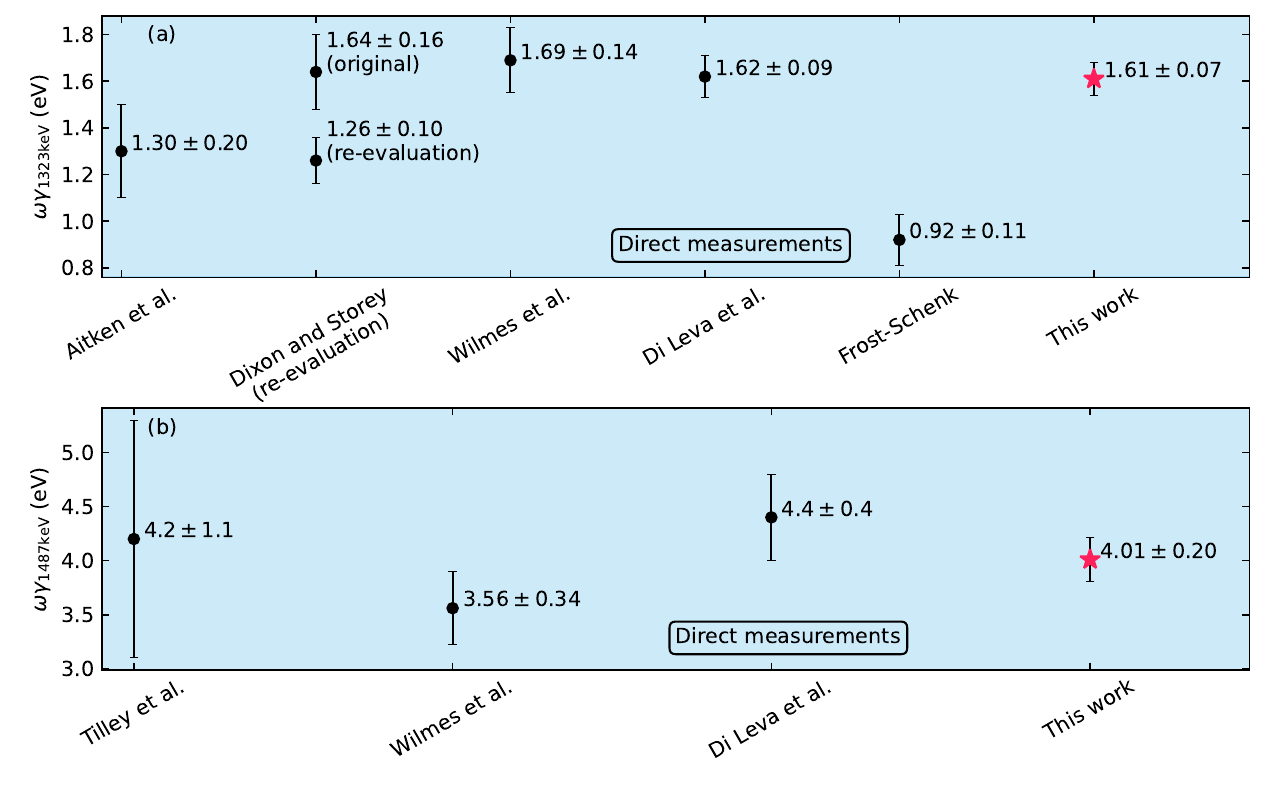}
    \caption{(a) Comparison $\omega \gamma(E_{\rm{c.m}} = 1323~keV)$ of this work and those of Refs.~\cite{Aitken_1970, Dixon_Storey_1971, Wilmes, 2017PhRvC..95d5803D, Frost-Schenk2020}. The re-evaluation of \citet{Dixon_Storey_1971} is discussed in Appendix \ref{sec:re_evaluation_of_wy_at_5337}. (b) Comparison of $\omega \gamma(E_{\rm{c.m}} = 1487~keV)$ from this work and those of Refs.~\cite{tilley_energy_1995,2017PhRvC..95d5803D, Wilmes}.}
    \label{fig:wy_comparison}
\end{figure*}
We determined the resonance energy, strength, and alpha width for the two resonances of interest and their values as shown in Figs.~\ref{fig:E_res_comparison} and \ref{fig:Gamma_comparison} and Table~\ref{table::res_strength_results}, respectively. We first discuss the resonance energy for both resonances studied here. For the lower energy resonance, the corresponding excitation energy of $5335.4 \pm 0.6$~keV, determined from this work, is compatible with the accepted value in the compilation \cite{tilley_energy_1995}, as well as those of \citet{Price57}, \citet{rogers_1972} and \citet{Frost-Schenk2020}. This suggests that the proposed value of $E_x = 5345.2 \pm 1.6$~keV in Ref.~\cite{2017PhRvC..95d5803D} is unlikely, thus further investigation is required. For the $E_x = 5501$~keV level, our result of $5493.2 \pm 0.6$~keV is 7.5~keV lower than the accepted value of $E_x = 5500.7 \pm 1.7$~keV \cite{tilley_energy_1995}. Contrary to most prior studies, our analysis is independent from the beam energy. In addition, as discussed in Sec.~ \ref{sec:5501 review}, the analysis of elastic scattering data~\cite{bardayan_2005, volya_2022} showed additional evidence that the corresponding level energy may be lower than the accepted value.

The strengths obtained from this work are compared with those from the literature as shown in Fig.~\ref{fig:wy_comparison}. For the $E_x = 5337$~keV level, our corresponding resonance strength of $1.61 \pm 0.07$~eV is compatible with the accepted value of $1.64 \pm 0.16$~eV \cite{tilley_energy_1995}, and most of the literature values. However, the re-evaluation of \citet{Dixon_Storey_1971}'s work (see Appendix~\ref{sec:re_evaluation_of_wy_at_5337}) using the updated resonance strengths for the reactions that were used for the relative measurements gives a much lower value of $1.26 \pm 0.10$~eV, and this is somewhat inconsistent with others in the literature (see Fig.~\ref{fig:wy_comparison}). Given the result of this work and those of \citet{Wilmes} and \citet{2017PhRvC..95d5803D}, the accepted value \cite{tilley_energy_1995} of $1.64 \pm 0.16$ is favored over the re-evaluated one. Nevertheless, as \citet{rogers_1972} and \citet{dixon_storey_1977} rely on the strength corresponding to the $E_x = 5337$~keV level to normalize the corresponding strengths of several higher energy levels in $^{19}$F, additional studies are needed. For the $E_x = 5501$~keV level's corresponding resonance strength, our result of $4.01 \pm 0.20$~eV is compatible with the two published values in \citet{Wilmes} and \citet{2017PhRvC..95d5803D}, and the accepted value of $4.2 \pm 1.1$~eV in the compilation~\cite{tilley_energy_1995}. 

Finally, the alpha width for the $E_x = 5337$ keV level determined from this work is compatible with those of Refs.~\cite{Price57, Wilmes, volya_2022, Goldberg2022}. However, the larger alpha width proposed by \citet{2017PhRvC..95d5803D} is incompatible with our work. For the $E_x = 5501$ keV level, our result confirms the larger alpha width suggested by \citet{2017PhRvC..95d5803D} and is compatible with that of \citet{Wilmes}, but is incompatible with those of Refs.~\cite{Price57, volya_2022, Goldberg2022}. The discrepancies of the alpha width for both levels require further investigations, as they have a large impact on the astrophysical reaction rates of the $^{15}$N$(\alpha, \gamma)^{19}$F reaction at temperatures relevant to low mass AGB stars, which we discuss in the following section.

\section{Astrophysical Reaction Rates} \label{sec:rr}
The total reaction rate for the $^{15}$N$(\alpha,\gamma)^{19}$F reaction was determined using the computer code \verb+RatesMC+ \cite{longland_ratesmc}, which calculates total reaction rate and uncertainty from resonant and nonresonant input parameters using the Monte Carlo method described in \citet{LONGLAND20101}. 

\begin{figure}
    \centering
    \includegraphics[width = \linewidth]{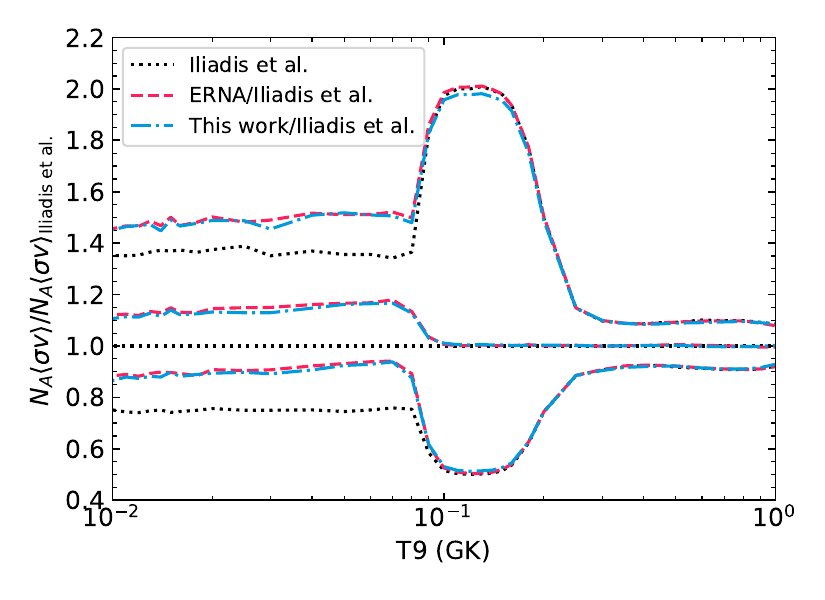}
    \caption{Ratio of the total reaction rates calculated from this work and \citet{2017PhRvC..95d5803D} to that of \citet{iliadis_charged-particle_2010}, including respective error bands. The ratio of this work is shown in blue dash-dotted lines, \citet{2017PhRvC..95d5803D} is shown in red dashed lines, and \citet{iliadis_charged-particle_2010} is shown in black dotted lines.}
    \label{fig:reaction_rate_comparison}
\end{figure}

Using the \verb+RatesMC+ input file provided in \cite{ILIADIS2010251} and modifying it for the two resonances using parameters found in this work and those of \citet{2017PhRvC..95d5803D}, two reaction rates are calculated. The comparison between those two rates and that of \citet{iliadis_charged-particle_2010} is presented in Fig.~\ref{fig:reaction_rate_comparison}.

For $T < 0.1$~GK, \citet{2017PhRvC..95d5803D} and \citet{buompane_recent_2022} suggested an increase of about 15\% to the median reaction rates due to the increased alpha widths for both resonances, while the uncertainty has been reduced by about 5\%. Both of these suggestions are confirmed by this work for the same reasons. In addition, our work has reduced the reaction rate upper limit by about 5\% at temperatures near $0.1$~GK. The impact of the updated reaction rates at temperatures relevant to AGB stars will be investigated and discussed elsewhere. 

\section{Summary and Conclusion} \label{sec:sum}
The resonances at $E_{\rm{c.m.}} = 1323$ and $1487$~keV in the $^{15}$N$(\alpha, \gamma)^{19}$F reaction have been studied using gamma spectroscopy. We determined the excitation energies of these two resonances. Our results are consistent with the accepted value of $1323 \pm 2$~keV for the lower energy resonance, but differ from the value of $1331.4 \pm 1.6$~keV reported in \cite{2017PhRvC..95d5803D}. For the higher energy resonance, our measurement indicates an energy that is 7.5~keV lower than the accepted value of $1487 \pm 1.7$~keV. This value is consistent with the elastic scattering measurements of \citet{smotrich_elastic_1961}, \citet{volya_2022}, and \citet{Goldberg2022}, but is not consistent with other radiative capture measurements. We recommend a value of $5493.2 \pm 0.6$~keV for the excitation energy, or a center of mass energy of $1479.4 \pm 0.6$~keV for this resonance.

Although the $E_{\rm{c.m.}} = 1323$~keV resonance strength in the compilation \cite{tilley_energy_1995} should be updated with those of the latest measurements, we found the original published value is compatible with our present work. The two resonance strengths studied in this work are consistent with most literature values with our results further reducing the uncertainties. An updated reaction rate is calculated using our results for the higher resonance energy, the alpha widths and resonance strengths. We confirmed the 15\% reaction rate increase at $T < 0.1$~GK from \cite{2017PhRvC..95d5803D}, a temperature range important for low mass AGB stars, indicating that the increased reaction rate impact to the $^{19}$F abundance in AGB stars needs further investigation.

\begin{acknowledgments}
We are extremely grateful to the technical staff at NSL. We also thank Antonino Di Leva for the useful discussion on the astrophysical reaction rate. This research utilized resources from the Notre Dame Center for Research Computing and is supported by the National Science Foundation (NSF) under Grants No. PHY-2011890 and PHY-2310059 (Nuclear Science Laboratory), and PHY-1430152 (JINA Center for the Evolution of the Elements).
\end{acknowledgments}

\appendix

\section{Re-evaluation of the $E_{\rm{c.m.}} = 1323$~keV ($E_{x} = 5337$~keV) resonance strength of \citet{Dixon_Storey_1971}}
\label{sec:re_evaluation_of_wy_at_5337}
In \citet{Dixon_Storey_1971}, three values have been reported for the $E_{\rm{c.m.}} = 1323$~keV resonance strength (see Sec.~\ref{sec:5501 review}). They reported a resonance strength relative to that of the yield of the $E_{\alpha} = 1532$~keV resonance in the $^{14}$N$(\alpha,\gamma)^{18}$F reaction given by \citet{parker_n_1968}. They first corrected a mistake in \citet{parker_n_1968} by properly treating the $E_{\alpha} = 1532$~keV resonance strength in the center-of-mass frame. Then, using updated stopping power tables, they concluded that the corrected resonance strength of the $E_{\alpha} = 1532$~keV resonance in the $^{14}$N$(\alpha,\gamma)^{18}$F reaction was $\omega \gamma (^{18}\rm{F; 1532~keV}) = 1.34 \pm 0.11$~eV \cite{Dixon_Storey_1971}. We re-evaluated this strength by comparing the yield of the resonances at $E_{\alpha} = 1140$~keV and $E_{\alpha} = 1532$~keV that were provided in \citet{parker_n_1968}. Using the more recent $E_{\alpha} = 1140$~keV resonance strength of \citet{gorres_2000} and modern stopping power tables~\cite{ziegler_srim_2010}, we calculated an updated resonance strength for the $E_{\alpha} = 1532$~keV resonance of $1.20 \pm 0.09$~eV by comparing the ratio of these two strengths
\begin{equation}
\begin{split}
    \frac{\omega \gamma (^{18}\rm{F; 1532~keV})}{\omega \gamma (^{18}\rm{F; 1140~keV})} &= \frac{\epsilon_r (1532)}{\epsilon_r (1140)}\frac{\lambda_r^2(1140)}{\lambda_r^2(1532)}\frac{Y(1532)}{Y(1140)}\\
    &= \frac{\epsilon_r(1532)}{\epsilon_r(1140)}\frac{E_r^{lab}(1532)}{E_r^{lab}(1140)}\frac{Y(1532)}{Y(1140)}.
\end{split}
\label{eq:res_strength_ratio}
\end{equation}
With this new value, we updated \cite{Dixon_Storey_1971}'s resonance strength for the $^{15}$N$(\alpha,\gamma)^{19}$F resonance to $1.46 \pm 0.17$~eV, about 10\% lower than the original published value.

\citet{Dixon_Storey_1971} also calculated the relative strength using the $E_{p} = 898$~keV resonance in the $^{15}$N$(p,\alpha_1 \gamma)^{12}$C reaction reported by \citet{gorodetzky_cascades_1968}, where \cite{Dixon_Storey_1971} calculated the resonance strength for the $E_{p} = 898$~keV resonance was $480 \pm 48$~eV. Using a more recent publication \cite{zijderhand_strong_1986} for the strength of the $E_{p} = 898$~keV resonance, we calculated a value of $293 \pm 38$~eV, which is almost 40\% lower than that of \citet{Dixon_Storey_1971}. \citet{zijderhand_strong_1986} report that their significant decrease in the $E_{p} = 898$~keV resonance strength likely comes from a background contribution from the $^{15}$N$(p,\alpha_0)^{12}$C reaction in which the $^{12}$C~(G.S.) recoils were not accounted for in the charged particle spectroscopy of~\cite{gorodetzky_cascades_1968}. With this updated value, the strength of the $^{15}$N$(\alpha,\gamma)^{19}$F reaction that corresponds to the $E_{x} = 5337$~keV level was reduced to $1.05 \pm 0.15$~eV. The primary source of uncertainty was from the uncertainty in the strength of the $(p, \alpha_1 \gamma)$ resonance. Following \citet{Dixon_Storey_1971}'s approach, the weighted average of the results of the two re-evaluated relative methods and the absolute resonance strength reported in their work yields $\omega \gamma = 1.26 \pm 0.10$~eV (see Fig.~\ref{fig:wy_comparison}), significantly lower than their original weighted average strength of $1.64 \pm 0.16$~eV.


%

\end{document}